# Energy effects on alloying through a re-interpretation of Miedema's parameters

T. Rajasekharan[1,*] and V. Seshubai[2,**]

[1]*Defence Metallurgical Research Laboratory, Kanchanbagh P.O. Hyderabad 500 058, India,*
[2]*School of Physics, University of Hyderabad, Hyderabad 500 019, India.*

Miedema's parameters $\phi$ and $N$ are quoted precise to the second decimal place and are considered very accurate after adjustment in their values to correctly predict the signs of the heats of formation of more than 500 metallic binary systems. In this paper, we argue that $\phi$ is proportional to the electronegativity ($\chi_M$) of metallic elements and that $N^{1/3}$ is the valence effective on a linear bond. We show that, with the above hypothesis, $\chi_M$ can be derived in the absolute sense, following Gordy, as the electrostatic potential due to the effective charge of the nucleus of an atom in a metal, felt at the mid-point of a bond to the nearest atom. The exponent $1/3$ has its origin in the high ligancy and resonance in metals. An excellent agreement with the hypothesis is observed for the hypoelectronic elements, while the buffer and hyperelectronic elements show interesting variations. It is argued that unlike atom pair bond energies, which are correlated to the distribution of binary systems on Miedema's ($|\Delta\phi|$, $|\Delta N^{1/3}|$) map, originate from the scatter in the distribution of elements in the electronegativity versus $N^{1/3}/R$ graph. Since it is the scatter that is important, we see that the small corrections effected by Miedema et al. to the parameters $\phi$ and $N$ are significant in deciding the energies involved. An interpretation is provided for the energy effects involved in the alloying of metals.

# I. Introduction

A study of the large amount of experimental data that exists in the literature on metallic alloy phase diagrams, using the average properties or lumped constants for atoms such as electronegativity and valence can be expected to help frame new laws. Such laws could be useful in predicting the outcome of experiments / processes in laboratories and industries, without resorting to extensive calculations every time. Ab initio theoretical efforts might also benefit from such a study by improving the necessary approximations.

Miedema's semi--empirical theory[1-4] for the heats of formation ($\Delta H$) of metallic alloys is reputed for its ability to predict *the signs of* $\Delta H$ with almost 100% accuracy; but its ability to predict accurately the numerical values has been disputed[5,6]. Miedema's equation for $\Delta H$, for a compound at equiatomic composition is given below:

$$\Delta H = \left[ -\Delta \phi^2 + \frac{Q}{P} \left( \Delta N^{1/3} \right)^2 - \frac{R}{P} \right] \qquad \ldots(1).$$

$\phi$ and N are called the work function and the 'electron density at the boundary of the Wigner-Seitz cells' of the elements. The numerical values of $\phi$ and N were obtained from Pauling's electronegativity and experimental bulk modulus of the elements, respectively. Their values were then adjusted within uncertainties in their determination to predict the signs of $\Delta H$ of more than 500 binary systems, using Eq. (1), with nearly 100% accuracy. Considerable amount of empiricism was involved in developing Miedema's equation. It was reported that the choice of $(\Delta N^{1/3})^2$ in Eq. (1), rather than $(\Delta N)^2$ which followed from Miedema's theoretical arguments, gave a better fit to the experimental data[2]. $P$ and $Q$ then become universal constants for all elemental combinations with $Q/P = 9.4$ Volts/(d.u.)$^{1/3}$, obtained from a fit of the experimental data. The term $R/P$, increasing with the valence of the $p$-metals, was introduced to correctly reproduce the empirically observed data on the signs of $\Delta H$ of transition metal - $p$-metal combinations. Miedema's equation is essentially empirical in nature with scope for reinterpretation of the physical meanings of the parameters ϕ and $N$.

The prediction of the crystal structures adopted by intermetallic compounds in binary phase diagrams has also interested scientists for a long time and they have worked essentially using structural maps with different pseudo-potential-derived or empirically derived parameters as coordinates[7-14]. Intermetallic compounds with different crystal structures were separated into different regions on such maps. Rajasekharan and Girgis studied[15] the behaviour of intermetallic compounds on structural maps with $\Delta \phi$ and $\Delta N^{1/3}$ as coordinates (*RG* maps), where $\Delta$ denotes the difference in the quantities between two alloying elements. Earlier attempts[16] to construct structural maps with the absolute values $|\Delta \phi|$ and $|\Delta N^{1/3}|$ as coordinates were unsuccessful. Rajasekharan and Girgis had shown that their maps could predict concomitant and mutually exclusive structure types in binary phase diagrams. This was a surprising result because Miedema's theory is supposed to be isotropic. Also, the structural energies are generally assumed to be small. Hence the observed systematics in the behaviour of binary systems on the ($\Delta \phi$, $\Delta N^{1/3}$) map is not anticipated from the current understanding of the alloying of metals. In a recent paper[17], we have shown that the above observations call for a reinterpretation of Eq. (1) as the energy of the nearest neighbour unlike atom–pair bond. It also follows that the bond remains identical in all the intermetallic compounds occurring in the same binary system at different compositions. A 'bond energy' can then be defined in alloys as for a conventional chemical bond[18] which remains more or less the same irrespective of the nature of the functional groups attached to the atoms on the bond.

Metallic elements do not have enough electrons to form conventional chemical bonds. Pauling had proposed[19-22] that resonance among various available bond positions can be a solution to this problem. He pointed out that properties of metallic elements such as melting points, boiling points, hardness etc. vary in a systematic way across the periodic table[19]. He attributed these variations to the variation in the number of covalent bonds formed between the atoms due to the variation in metallic valence. He had defined a set of valence for metallic elements from the above observations[23]. He could explain[19] qualitatively many properties of transition metals such as inter-atomic distances, characteristic temperatures, hardness, compressibility, coefficient of thermal expansion; and the atomic saturation moments of the ferromagnetic elements Fe, Co and Ni and their alloys. However, the difficulties in making Resonating Valence Bond calculations and also the widespread belief from long time ago[24] that metals cannot be covalently bonded have prevented extensive use of the theory. Mohallem et al.[25] have reported a fully ab initio valence bond calculation for lithium clusters which confirms the importance of the metallic orbital and the covalent character of the metal-metal bond.

As concluded from the observations on the Rajasekharan-Girgis maps[17], if Eq. (1) were to represent the energy of the unlike-atom-pair bond, the form of the first term immediately suggests its origin as due to the electronegativity difference between the atoms. A compound forms by the breakage of A−A and B−B bonds and the formation of A−B bonds. The bond formed between A and B will have a certain ionic character and the increase in stability of the heteronuclear bond A−B in comparison with the average of the homonuclear bonds can be attributed to the ionicity in the bonds. The negative contribution to the bond energy due to this effect would be equal to $-(\Delta\chi)^2$ eV/bond[26, 27].

## II. Electronegativity and Valence of metals

We recall that the values of Miedema's parameter $\phi$ were obtained by scaling up Pauling's electronegativity values ($\chi$) and then adjusting them within the uncertainty in their determination. Fig.1 shows a plot between $\phi$ and $\chi$ for elements.

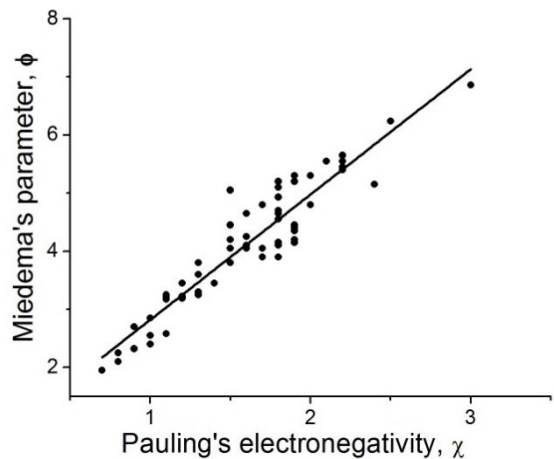

**Fig. 1** A plot of Miedema's parameter $\phi$ versus Pauling's electronegativity values $\chi$ for elements. The equation for the best fit line is $\chi = (\phi - 0.662)/ 2.157$.

We obtain electronegativity values *for metals* $\chi_M$ by scaling down the $\phi$ values while retaining the advantages of the corrections made, by using the linear relationship between the two, as given below:

$\chi_M = (\phi - 0.662)/ 2.157.$

As mentioned earlier, Pauling had arrived at the numerical values for valences of metals by studying the variation of their physical properties such as melting points, hardness etc. across the periods in the periodic table[19]. We see from Fig. 2 (a) that in the first long period of the periodic table, Miedema's $N$ are comparable in magnitude, and vary in the same way with the atomic number of elements, as do Pauling's valences. From Fig. 2 (b), we see that *N* follows the variation in the melting points of elements, including the anomalous behaviour of Mn, more closely than Pauling's valences.

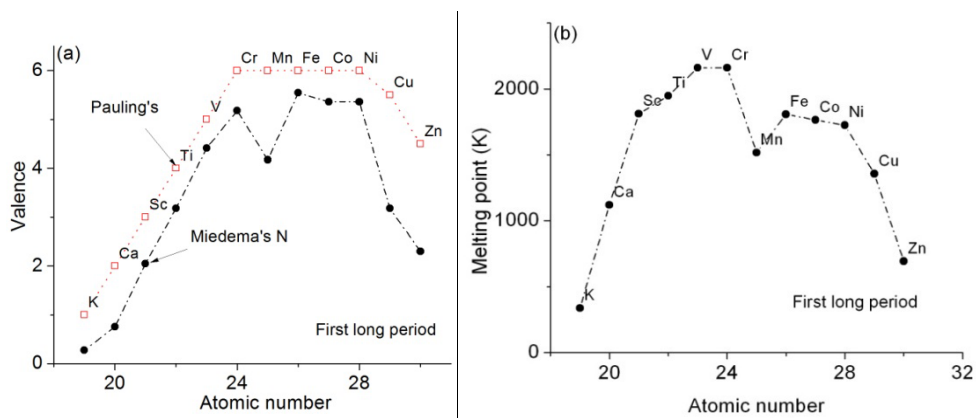

**FIG. 2. (a)** (color online). We see that, in the first long period of the periodic table, Miedema's *N* are comparable in magnitude, and vary in the same way with the atomic number of the elements, as do Pauling's valences. In **(b),** we see that *N* follows the variation in the melting and boiling points of elements, including the anomalous behaviour of Mn, more closely than Pauling's valences. The numerical values of N can be adjusted by a constant value to match Pauling's valences even better, but we have not attempted it herein since only the difference in $N^{1/3}$ of the elements are relevant to our (or Miedema's) conclusions.

Similar dependencies can be demonstrated for elements of the second (Figs. 3(a) and (b)) and third (Figs. 4(a) and (b)) long periods of the periodic table as well. We recall that *N* values, which are close to Pauling's valences for metals, had also been adjusted to predict the signs of *ΔH* accurately. These observations suggest the possibility that the corrected *N* are the accurate valences of metallic elements.

Metallic alloys are characterised[28] by high symmetry, large connectivity and high ligancy (usually ≥ 8). There are thus a large number of equivalent nearest neighbour unlike atom-pair bonds. The covalent bonds can be considered to resonate among the available equivalent positions. Since metals are electron deficient, there has to be a hypothesis on the factor of the valence electrons of the atoms that would be effective on a single bond. If we assume that Miedema's corrected electron density parameters $N$ can be accepted as the accurate valence of metals for reasons discussed earlier, we might assume that $N^{1/3}$ is a good approximation for the valence of a metal effective on a linear bond due to the high symmetry and resonance among a large number of equivalent bond positions. We note that Miedema's equation which accurately predicts the signs of *ΔH*, and also the Rajasekharan-Girgis lines which accurately predict concomitant structures in phase diagrams use the parameter $N^{1/3}$, and not *N*. An expression has been derived elsewhere[29] for the energy of the atom-pair bond which assumes $N^{1/3}$ to be the valence of a metal effective on a bond. The expression predicts the signs of the heats of formation of metallic alloys accurately and generates Miedema's constant $\sqrt{Q/P}$ without any empiricism.

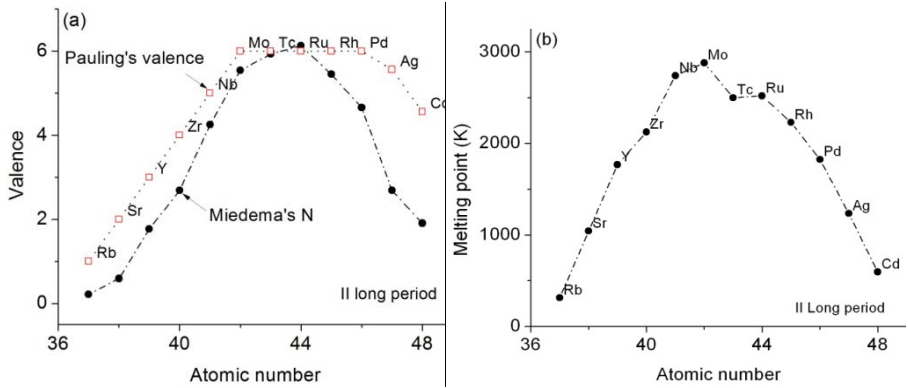

**FIG. 3. (a)** (color online). We see that, in the second long period of the periodic table, Miedema's $N$ are comparable in magnitude, and vary in the same way with the atomic number of elements, as do Pauling's valences. In **(b)**, we see that $N$ follows the variation in the melting points of elements more closely than Pauling's valences.

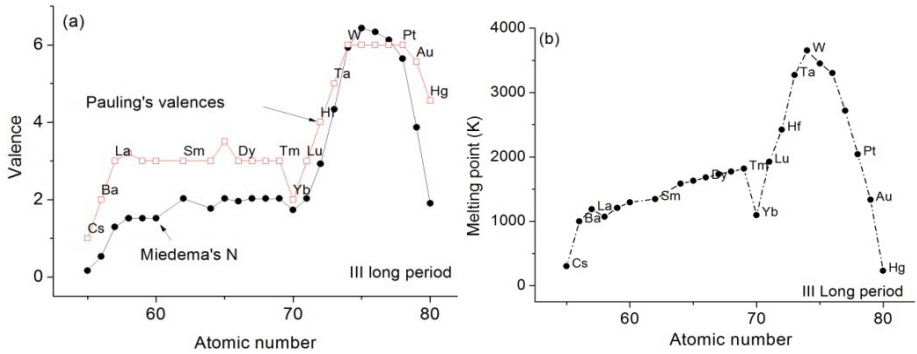

**FIG. 4. (a)** (color online). We see that, in the third long period of the periodic table, Miedema's $N$ are comparable in magnitude, and vary in the same way with the atomic number of elements, as do Pauling's valences. In **(b)**, we see that $N$ follows the variation in the melting points of elements more closely than Pauling's valences.

We show below that the electronegativity $\chi_M$ of metals as defined earlier, can be derived from Miedema's *N* values following Gordy's[30] approach. Electronegativity can be defined as the potential $(Z_{eff})e/R$, where $(Z_{eff})e$ is the effective nuclear charge of an atom felt by a valence electron at a distance *R* from its nucleus. To arrive at $Z_{eff}$, Gordy had proposed[30] that the electrons in the inner shells of an atom exert their full screening power while the valence electrons would exert only 50% of their screening power. This involves an assumption that 50% of the time, the valence electrons reside at a distance closer than R from the nucleus. Gordy observes that Pauling had used 40% screening power for the valence electrons and that the conclusions drawn are not very sensitive to small variations in the screening power assumed. As argued before, we assume that due to resonance and high symmetry, $N^{1/3}$ electrons would be effective as valence electrons on a linear bond on a time-average, and would be located at the midpoint of the bond with the next atom. The remaining $(N - N^{1/3})$ valence electrons stay closer to the nucleus on an average with spherically symmetric density, exerting their full screening power. The above assumption leads to the conclusion that $Z_{eff} = N^{1/3}e$. The electronegativity $\chi_M$ would then become

$$\chi_M = k\frac{(Z_{eff})e}{R} = k(N^{1/3})e/R, \text{ where k is a constant.}$$

In Fig. 5, we show a plot of $\chi_M$ versus $(N^{1/3})/R$ for metallic elements with $R$ as the metallic radius.

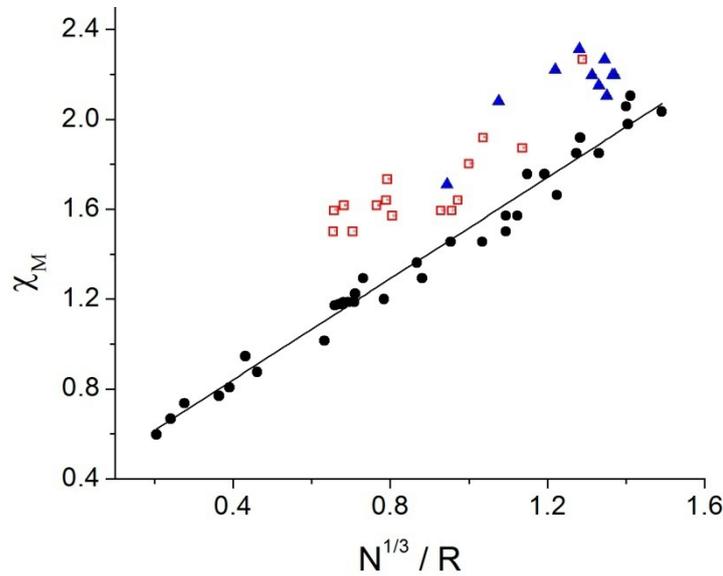

**Fig. 5.** (color online). A plot of $\chi_M$ versus $(N^{1/3})/R$ for metallic elements with $R$ as the metallic radii. A good linearity with a regression factor of 0.98 is observed for the hypo-electronic transition metals, s-elements and rare earths (closed circles, black online) supporting the hypothesis that $N^{1/3}$ can be considered the valence of the elements effective on a linear bond, and $\chi_M$ the electronegativity. The p-metals which are hyper-electronic (seen as unfilled squares, red online) and the buffer elements Ag, Au, Re, Ru, Os, Tc, Rh, Ir, Pt, and Pd (triangles, blue online) form separate groups.

Pauling had classified[39] elements as hyperelectronic, buffer and hypoelectronic based on their electronic structure and alloying behaviour. An exact linear correlation is observed in Fig. 5 for hypo-metals which include s-elements, most of the transition metals and rare earths, with a regression factor of 0.98. The hyperelectronic p-metals are separated based approximately on their valence, with the valence increasing as one moves away from the hypo-metal line. In Gordy's plot[30], the elements Ag and Au were exceptions and the elements Re, Ru, Os, Tc, Rh, Ir, Pt, and Pd were not considered. Those elements are classified as buffer elements by Pauling[39] and we note that they are separated from the hypo- and hyper- electronic elements in the figure. The increase in $\chi_M$ in the sequence of hypo- to buffer to hyper- electronic elements can be attributed to the decrease in radius $R$ due to an increase in valence. The excellent correlation in Fig. 5 shows that the electronegativities of the hypo-metallic elements can be calculated accurately from their $N^{1/3}$ values in the absolute sense.

In Fig. 6, we show a plot of Pauling's electronegativity values (i.e. $\chi$, before Miedema's correction) versus $N^{1/3}/R$. The larger scatter of points in Fig. 6 compared to that in Fig. 5 shows that the corrections made by Miedema et al. to their parameters have great significance. The corrections introduced from empirical heat of formation data had the effect of improving the linear correlation in Fig. 5, reinforcing the new interpretation of $\chi_M$ and $N^{1/3}$ introduced in this paper.

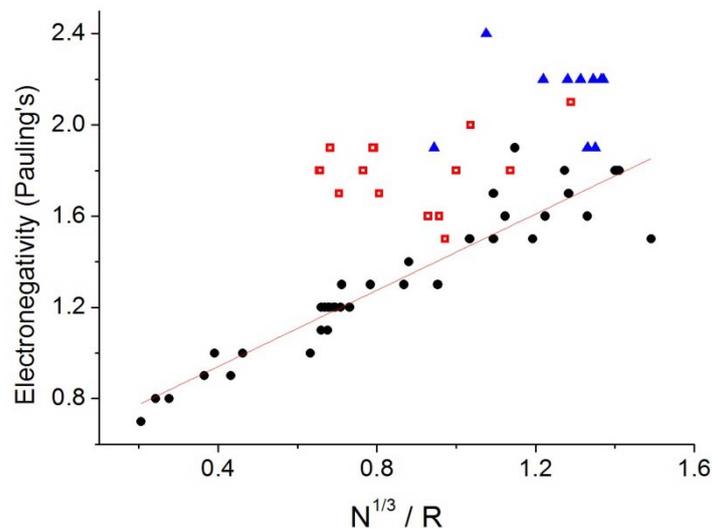

**FIG. 6** (color online). Plot of uncorrected Pauling's valence $\chi$ versus $N^{1/3}/R$ for metallic elements. The hypo-electronic transition metals, s-elements and rare earths (closed circles, black online), the p-metals which are hyperelectronic (unfilled squares, red online) and the buffer elements Ag, Au, Re, Ru, Os, Tc, Rh, Ir, Pt, and Pd (triangles, blue online) can be seen in the figure. The distribution of points is similar to that in Fig. 6, but the scatter in the points is more than in Fig. 5. It is interesting to note that the effect of the corrections implemented by Miedema et al. to their parameters using thermo- chemical data has been to refine the correlation between electronegativity and $(N^{1/3})/R$ in Fig. 5.

## III. The energy effects on alloying of metals

One of the puzzling aspects of Miedema's Eq. (1) has been that the best fit line between $\phi$ and $N^{1/3}$ of the elements has nearly the same slope as that of the line (of slope $\sqrt{Q/P}$) separating the compounds with negative and positive heats of formation on a $(|\Delta\phi|, |\Delta N^{1/3}|)$ map. In Fig. 7, we show a plot of $\phi$ versus $N^{1/3}$ for elements along with its best fit line. The best fit line can be seen to be located close to the line of slope $\sqrt{Q/P}$ = 3.07 passing through the origin, also shown in the same figure.

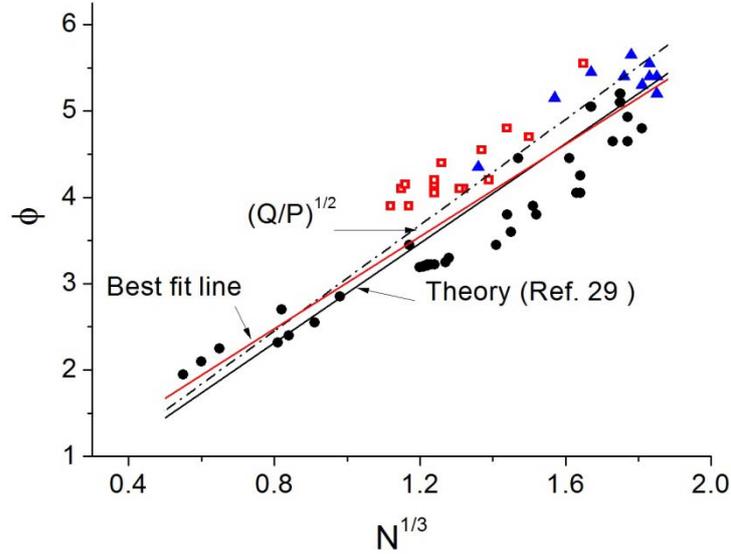

**FIG. 7** (color online). A plot of Miedema's parameter $\phi$ versus $N^{1/3}$ for metallic elements. The best fit line, a line of slope $\sqrt{Q/P}$ = 3.07 and another with the slope predicted by the theory in reference 29 are shown. The slopes are nearly the same for all the lines.

Let us imagine a number of straight lines parallel to the line of slope $\sqrt{Q/P}$ in Fig. 7. When two elements lie on any one of such lines, $\Delta\phi = \sqrt{Q/P}\,(\Delta N^{1/3})$ for their compound which will lie on the line of slope $\sqrt{Q/P}$, passing through the origin, on a $(|\Delta\phi|, |\Delta N^{1/3}|)$ map. The heat of formation of the compound would be zero according to Eq. (1). It is when compounds are formed between elements lying on two different such lines that the heat of formation of their compound differs from zero.

If electronegativity of all the metals were proportional to $N^{1/3}$ with the same slope in Fig. 7, the energy of all the compounds would have been identically equal to zero. The reason for non-zero bond energies is the scatter in the figure. Since the scatter is important, we see that the small corrections effected by Miedema et al. to the parameters $\phi$ and $N$ are significant in deciding the energies involved. Those corrections are also important in defining the electronegativity and valence values for metals. The scatter in Fig. 7 is similar to that in Fig. 5. In Fig. 5, electronegativity is more for some elements for the same $N^{1/3}/R$ value because of higher bond order and smaller radius. We see that the lines parallel to the $\sqrt{Q/P}$ line in Fig. 7 are arranged as per the valence from hypoelectronic to buffer to hyperelectronic elements. If the lines on which two atoms are located are far, then their electronegativity difference will be large for a given difference in $N^{1/3}$ and the lowering of energy on alloying would be more. Such effects would increase in the order of hypo-hypo, to buffer – hypo, to hyper—hypoelectronic elements, and would be minimal between atoms of the same group in the periodic table.

A brief description of the nature of the metallic bond in an alloy as discussed in Ref. 29 is in place here. In molecular compounds, the stoichiometry of the molecule and the molecular conformation are adjusted so that there are no charges of the same sign in the neighbourhood. In a metallic solid the situation is different. According to the conclusions drawn from the $RG$ lines, the metallic bond between two given unlike atoms is nearly identical at all compositions in the phase diagram.

It is different from the conventional chemical bond in the sense that there is an actual transfer of charge from one atom to the other to meet the requirement of electroneutrality. The atoms are thus partially ionised with a change in their sizes. There is an energy cost for this charge transfer. It is given[29] by $1.34\,\Delta\chi\,\Delta N^{1/3}$. Compared to this, if the lowering of the energy during bond formation due to electronegativity difference between the atoms $(-(\Delta\chi)^2)$, is more in magnitude, the bond will be stabilised. According to the Eq. (4) in ref. 29, the bond will be stabilised when $\Delta\chi_M/\Delta N^{1/3} > 1.34$. In a $(\Delta\phi, \Delta N^{1/3})$ map, this condition would turn out to be $\Delta\phi/\Delta N^{1/3} > 2.89$. The line corresponding to that condition is also shown in Fig. 7 and is located close to the $\sqrt{Q/P}$ line. Thus the magnitude of Miedema's constant $\sqrt{Q/P}$ is connected to the relationship between electronegativity and valence on the metallic bond and the magnitudes of bond energies stem from the scatter in Fig. 7 and hence in Fig. 5.

## ACKNOWLEDGEMENTS

TR thanks DMRL, Hyderabad, India for permission to publish this paper. VS thanks UGC, India for research funding under CAS program and CSIR, India for a research project.